\def\year{2020}\relax
\documentclass[letterpaper]{article} 
\usepackage{aaai20}  
\usepackage{times}  
\usepackage{helvet} 
\usepackage{courier}  
\usepackage[hyphens]{url}  
\usepackage{graphicx} 
\usepackage{graphicx}  
\usepackage{amsmath}
\usepackage{subcaption}
\usepackage{amsfonts}
\usepackage{tikz}

\title{Disentangling Speech and Non-Speech Components for Building Robust Acoustic Models from Found Data}

\author{Nishant Gurunath \textsuperscript{\rm 1}, Sai Krishna Rallabandi \textsuperscript{\rm 1}, Alan Black \textsuperscript{\rm 1} \\ \textsuperscript{\rm 1} Carnegie Mellon University \\ 
5000 Forbes Ave \\
Pittsburgh, Pennsylvania 15213 \\
ngurunath@cmu.edu 
}
\begin{document}

\maketitle

\begin{abstract}
In order to build language technologies for majority of the languages, it is important to leverage the resources available in public domain on the internet - commonly referred to as `Found Data'. However, such data is characterized by the presence of non-standard, non-trivial variations. For instance, speech resources found on the internet have non-speech content, such as music. Therefore, speech recognition and speech synthesis models need to be robust to such variations. In this work, we present an analysis to show that it is important to disentangle the latent causal factors of variation in the original data to accomplish these tasks. Based on this, we present approaches to disentangle such variations from the data using Latent Stochastic Models. Specifically, we present a method to split the latent prior space into continuous representations of dominant speech modes present in the magnitude spectra of audio signals. We propose a completely unsupervised approach using multinode latent space variational autoencoders (VAE). We show that the constraints on the latent space of a VAE can be in-fact used to separate speech and music, independent of the language of the speech. This paper also analytically presents the requirement on the number of latent variables for the task based on distribution of the speech data.
\end{abstract}

\noindent Speech synthesis has taken some major strides in past few years especially in the form of text-2-speech synthesis (TTS) models. However, most of the work that has been carried out involves carefully recorded speech data. Generation of such vast amount of data for every application is a daunting task. On the other hand, there is a plethora of speech data that is available on the internet such as news broadcasts, press conferences, audio books etc - also referred to as \textit{Found Data}. The only hindrance in utilizing such data for speech based machine learning models is that this found data is characterized by noise or music in the background. Presence of noise / music degrades the performance of such models. One of the solutions to this problem is source separation - separating out speech from music in the audio. There have been several attempts to accomplish this task using both classical speech processing techniques as well as deep learning models.\\

\cite{6287816} proposed a matrix factorization of the magnitude spectrogram of audio that utilizes the periodicity in music and sparseness in speech to separate the two. However, this technique requires a lot of hyperparameter tuning depending on the type of background music and also degrades the quality of separated speech to some extent. REPET \cite{6269059} also involves music separation by exploiting its periodic nature but on occasions still leaves a residual music in the background. Most of the work in source separation using deep learning has been supervised \cite{SVSGAN}, \cite{Disc}, \cite{TFGAN}, \cite{spen}, i.e. they had both noisy and clean versions of the data. However most of the times, especially with found data, we don't have the clean version of the data. \\

There has also been some focus on source separation using unsupervised models. \cite{Hsu2018DisentanglingCS} takes the approach of data augmentation by adding different background noise to the clean data and then training an adversarial classifier to make these augmented versions of data indistinguishable from the original speech. However, this method again requires a clean version of data first and additional data augmentation that is representative of the noise in the background. Therefore essentially, this is a semi-supervised approach that requires labels for clean and noisy data. One other semi-supervised approach is using domain adaptation \cite{domadp} where output is made to follow the clean data domain while making the encoding for clean and noisy data domain indistinguishable using a adversarial classifier. However, this approach requires speech content in both clean and noisy version of data to be very similar for domain adaptation to occur. \\

We propose a completely unsupervised approach using multinode variational autoencoders (VAE) combined with robust principal component analysis (RPCA) \cite{6287816} as a post-processing step. Our goal is to enable the use of found data for downstream TTS applications. Therefore, the data we target is predominantly speech with music in the background. We apply this approach on two datasets:- Wilderness\cite{wildernessdataset} and Hub4. Wilderness consists of Bible recordings in 699 languages while Hub4 is a news broadcast dataset in English. Both of these datasets contain music/noise in the background. We show that the proposed approach separates out the dominant mode, speech, from a noisy audio and improves the performance of the downstream tasks irrespective of the language of the speech. \\

This paper is organized as follows: section 1 discusses the variational autoencoder framework, section 2 talks about source separation using VAE, section 3 addresses the extension to multinode VAE architeture and section 4 discusses post-processing using robust principal component analysis (RPCA). Section 5 analyses the source separation capacity and architectural requirements of the proposed model. Section 6 reports the performance of the proposed model for source separation and for downstream TTS applications. We conclude in section 7.

\section{Variational Autoencoder}
Variational autoencoder model in this paper follows the standard formulation consisting of an inference network with a speech encoder $p(z|x)$ and a latent space decoder $p(x|z)$, where $x$ and $z$ represent the input and the latent space random variables respectively. 

\begin{figure}[h!]
    \centering
    \includegraphics[scale=0.4]{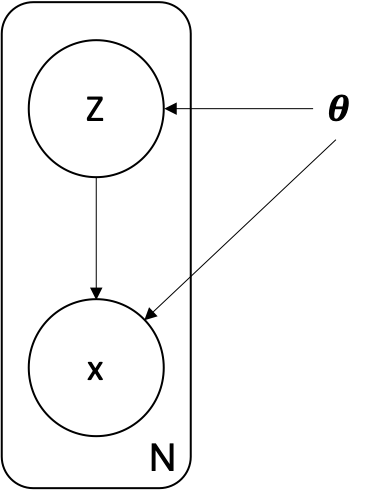}
    \caption{Latent Variable Model - Variational Autoencoder}
    \label{fig:vae_graph}
\end{figure}

The figure \ref{fig:vae_graph} depicts the latent variable model for variational autoencoder. The true posterior density is intractable.

\begin{equation}
    p(z|x) = \frac{p(x|z)p(z)}{p(x)}
\end{equation}

We then approximate the true posterior $p(z|x)$ with a variational distribution $q(z|x)$ that has a prior $p(z)$. The objective can be represented by the evidence lower bound (ELBO) or variational lower bound on the likelihood of the data. 
\begin{equation}
    \log{p(x)} \geq \mathcal{L}(x)
\end{equation}

\begin{equation}
\mathcal{L}(x) = \mathbb{E}_q[\log(p(x|z))] - D_{KL}(q(z|x) || p(z)) 
\label{Likelihood}
\end{equation}

where $\mathcal{L}(x)$ denotes the variational lower bound on the likelihood of the data and $D_{KL}$ is the Kullback-Leibler divergence. We write the first term as a mean squared error (MSE) between the reconstructed and the original data and the prior p(z) follows a standard normal distribution $\mathcal{N}(0,I)$.

\section{VAE for Source Separation}
As shown in the previous section, a variational autoencoder reconstructs the input data conditioned on the latent space. The latent space is constrained to follow a certain prior distribution, such as Gaussian distribution. \cite{JMLR:v19:17-704} shows that this formulation is equivalent to minimizing the alternative lower bound function 

\begin{equation}
\begin{aligned}
minimize \;\; & n \cdot rank[\, L \,] \,+ \parallel S \parallel_0   \\
M & = L + S \\
\end{aligned}
\end{equation}

where M is the original data matrix, L is a low-rank matrix and S is a sparse matrix and $\parallel.\parallel_0$ denotes the $l_0$ norm. This is shown to be equivalent to an RPCA problem if an optimum solution exists otherwise it's known to smooth out undesirable erratic peaks from the energy curve. \cite{JMLR:v19:17-704} also presents some interesting results on VAE and it's separation properties. We rewrite some of the results what it means in our context here.

\begin{itemize}
    \item This formulation of variational autoencoders is shown to perform robust outlier removal in the context of learning inlier points constrained to a manifold of unknown dimension. In simple terms this means, VAE has the property to remove sparse components in the input data distribution and accordingly reduces the latent space to a required (unknown) dimension.
    \item VAE also help smooth out undesirable minima from the energy landscape of the optimization problem which differentiates it from traditional deterministic autoencoders.
\end{itemize}
 
Since our goal is to enhance speech synthesis and speech recognition performances on the 'found' data, we target audio data that is predominantly speech with some music (almost uniform) in the background, for instance, news broadcasts and audio books. We will later show that the presence of the background music can effect the speech synthesis performance drastically. It's been shown that speech and music distributions in audio are quite distinct \cite{SM_diff} \cite{SM_diff1}. As a result VAE has the tendency to remove the sparse outlier - music from the audio. \\ 

In case of multiple speakers in the input audio there can be multiple modes in the speech distribution as well. This can be solved by have multiple nodes in the latent space. This is possible because all latent variables are initialized at random and pick multiple speech modes from the distribution. Later in the results section, we are going to analyze this and the requirement on the number of nodes in the latent space depending on the input data distribution. We also talk about how the performance of the output speech changes based on the intensity/loudness of the music in the background.

\section{Multi-node VAE Model Architecture}

\begin{figure}
    \centering
    \includegraphics[scale=0.31]{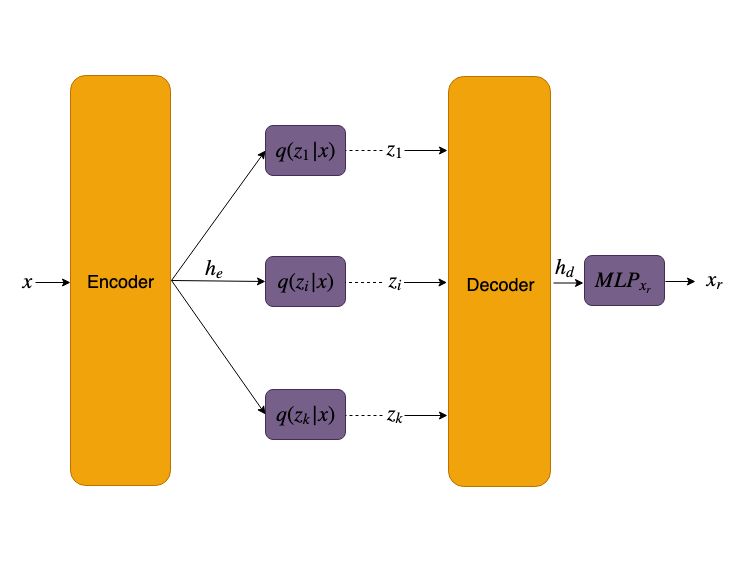}
    \caption{Multi-node VAE model. Dashed lines represent sampling using reparametrization. Encoder and Decoder are Bi-LSTM networks. Purple blocks are fully connected layers.}
    \label{fig:VAE}
\end{figure}

\begin{figure}[h!]
    \centering
    \includegraphics[scale=0.4]{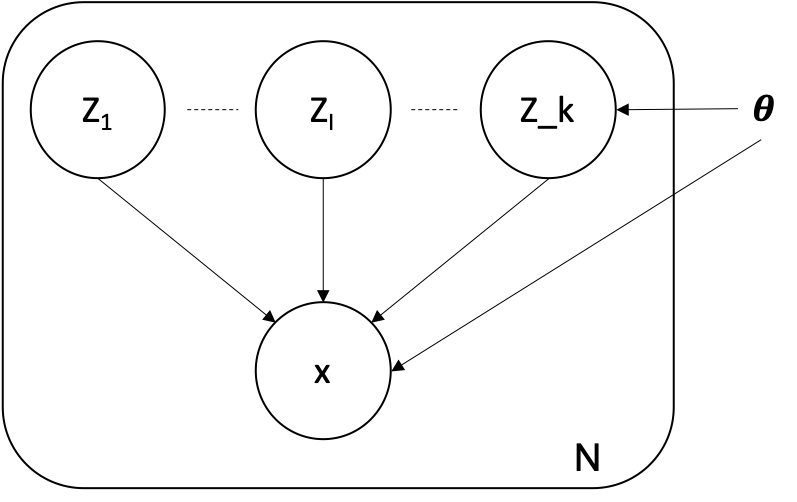}
    \caption{Latent Variable Model - Multinode Variational Autoencoder}
    \label{fig:multinodevae_graph}
\end{figure}

The Figure \ref{fig:VAE} depicts the multi-node variational autoencoder architecture. It consists of an Bi-LSTM encoder ($LSTM_E$) for the inference network that captures latent space distributions $p(z_1|x), p(z_2|x), \dots, p(z_k|x)$ where $x$ is the magnitude spectrogram of the input audio, $z_1, z_2, \dots, z_k$ are the latent variables and $k$ is the number of latent variables. The reconstruction network is a Bi-LSTM decoder ($LSTM_D$) which generates the reconstructed input distribution at each time-step $p(x_r^t|x_r^{t-1},z_1,z_2,\dots,z_k)$ conditioned on the reconstructed input from the previous time-step and the latent space. 

\begin{equation}
h_e^t, c_e^t  = LSTM_E(x^t,h_e^{t-1},c_e^{t-1})
\end{equation}
\begin{equation}
\mu_i^t = MLP_{\mu_i}(h_e^t) \; \forall i \in {0,\dots,k}
\end{equation}
\begin{equation}
logvar_i^t = MLP_{\sigma_i}(h_e^t) \; \forall i \in {0,\dots,k}
\end{equation}
\begin{equation}
q(z_i|x) = \mathcal{N}(\mu_i,\exp(logvar_i))
\end{equation}
\begin{equation}
h_{d}^t, c_{d}^t = LSTM_D(\phi^t, z_1^t, \dots, z_k^t,h_d^{t-1},c_d^{t-1})
\end{equation}
\begin{equation}
\phi^t = MLP_{\phi}(h_d^{t-1})
\end{equation}
\begin{equation}
x_r^t = MLP_{x_r}(h_d^t)
\end{equation}

where Bi-LSTM refers to bidirectional long short term memory recurrent neural network, MLP refers to a multi-layer perceptron network, $h_e, c_e$ represent hidden and cell states of the encoder LSTM, $h_d, c_d$ represent hidden and cell states of the decoder LSTM and $\phi$ represents the context from the previous time-step of the decoder. The initial hidden state and the cell state of the decoder LSTM are learnable parameters. The latent variable model for the multinode variational autoencoder is shown in figure \ref{fig:multinodevae_graph}\\

The modified learning objective for a multinode VAE can be represented as an extension of equation \ref{Likelihood} as:

\begin{equation}
\begin{aligned}
\mathcal{L}(x) \ge & \; \mathbb{E}_q[\log(p(x|z_1,z_2,\dots,z_k))] \\& -\sum_{i=1}^k D_{KL}(q(z_i|x) || p(z_i)) 
\end{aligned}
\end{equation}

\section{Speech Enhancement}

The output of the VAE network from the above formulation removes the music from the audio however, replaces the music content with random noise instead of silence. There can be multiple post processing or speech enhancement techniques used to eliminate this residual noise such as speech enhancement neural networks or classical speech processing methods. Here in this paper we use robust principal component analysis (RPCA) \cite{6287816} to eliminate the background noise as it gives control over the quality of speech versus the amount of background noise. We follow the original formulation from the paper by expressing the speech separation as a matrix factorization problem. It represents the magnitude spectrogram of the audio signal as a sum of low rank matrix and a sparse matrix. The assumption here is that non-speech component (background noise) is low rank while the speech component is sparse. 

\begin{equation}
\begin{aligned}
minimize & \parallel L \parallel_* + \lambda \parallel S \parallel_1   \\
M & = L + S
\end{aligned}
\end{equation}

where $M \in \mathbb{R}^{n_1 \times n_2}$ is the magnitude spectrogram of the VAE output, $L \in \mathbb{R}^{n_1 \times n_2}$ is a low rank matrix, $S \in \mathbb{R}^{n_1 \times n_2}$ is a sparse matrix, $\parallel . \parallel_*$ is the nuclear norm and $\parallel . \parallel_1$ is the $L_1$ norm. $\lambda > 0$ is a hyperparameter that controls the rank and sparsity of $L$ and $S$ respectively. It is recommended in \cite{6287816} to use $\lambda = 1/\sqrt{max(n_1,n_2)}$ to obtain the best result. However, we only need to enhance the audio a little while retaining the speech quality so we use $\lambda = 0.3/\sqrt{max(n_1,n_2)}$. Instead of the hard mask in \cite{6287816} we used a soft mask as it resulted in a better quality and a smoother speech. The idea is to have a high value for the speech mask where the magnitude of the speech component is much greater the magnitude of the non-speech component.

\begin{equation}
|S|  > g|L|
\end{equation}

\begin{equation}
|S|^2  > g^2 |L|^2
\end{equation}

\begin{equation}
|M|^2  = |S|^2 + |L|^2
\end{equation}

\begin{equation}
|S|^2  > g^2 |M|^2 - g^2|S|^2
\end{equation}

\begin{equation}
|S|^2  > \frac{g^2}{1 + g^2} |M|^2 
\end{equation}

\begin{equation}
|S|  > \sqrt{\frac{g^2}{1 + g^2}} |M|
\end{equation}

\begin{equation}
\frac{|S|}{|M|} - \sqrt{\frac{g^2}{1 + g^2}} > 0 
\end{equation}

where $g \ge 0$ is the gain factor. We came up with a Sigmoid looking threshold for the mask which is still close to the hard mask but results in smoother speech transitions.  

\begin{equation}
W = \frac{1}{1 + \exp{(-\alpha(\frac{|S|}{|M|} - \sqrt{\frac{g^2}{1 + g^2}}))}}
\end{equation}

where $W \in \mathbb{R}^{n_1 \times n_2}$ represents the speech mask and the obtained speech spectrogram is

\begin{equation}
X_{speech}(i,j) = W(i,j)M(i,j) \; \forall i,j
\end{equation}

\section{Experiments}
We applied the multinode VAE model on two datasets:- Wilderness and Hub4. Wilderness dataset consists of Bible recordings in 699 languages with music in the background. We carried out full experiments on two languages:- Dhopadhola (an African language) and Marathi (an Indian language). The results presented here are based on the model that was trained on languages different than the ones that are reported/tested. Hub4 consists of news broadcast recordings in English with various forms of noise in the background, such as music, clapping, roaring etc. We used about 2 hrs of training data for both datasets consisting 1 hr of speech only data and 1 hr of speech-music data. \\

For these experiments, the VAE model consists of input magnitude spectrogram of dimension 512, Bi-LSTM encoder and decoder with hidden size of 512, each of the fully connected layers for latent variables and decoder context from the previous time-step of dimension 64 and the final output layer with a dimension same as the input dimension. We trained for 50 epochs with annealing weight for KL-Divergence loss, this is explained in detail below. For both datasets, we used an ADAM optimizer with a learning rate of 1e-3. \\

\begin{figure}[h!]
    \centering
    \begin{subfigure}{0.2\textwidth}
    \includegraphics[scale=0.55]{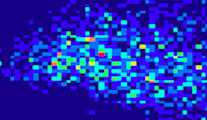}
    \caption{Wilderness}
    \end{subfigure}
    \hspace*{\fill}
    \begin{subfigure}{0.2\textwidth}
    \includegraphics[scale=0.55]{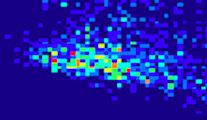}
    \caption{Hub4}
    \end{subfigure}
    \caption{Input Data Distributions for (a) Wilderness (b) Hub4. The red dots show the high density regions in each distribution.}
    \label{fig:distributions}
\end{figure}

The input data distributions for the two dataset are shown in Figure \ref{fig:distributions}. These 2-dimensional distributions are obtained after applying PCA to the magnitude spectrogram and plotting the histogram of the first two components. This figure shows the high density regions of the two distributions. As we can observe, the wilderness distribution has one significant high density region while the hub4 distribution consists of multiple high density regions. Hence, Hub4 data will have more dominant speech modes than the wilderness data. This is probably because there are multiple speakers in Hub4 as well as news broadcast speech has more variance as compared to bible recordings. This gives us an approximate idea that Hub4 multinode VAE model will require more nodes in the latent space than the model for Wilderness dataset. \\

\begin{figure}[h!]
    \centering
    \includegraphics[scale=0.37]{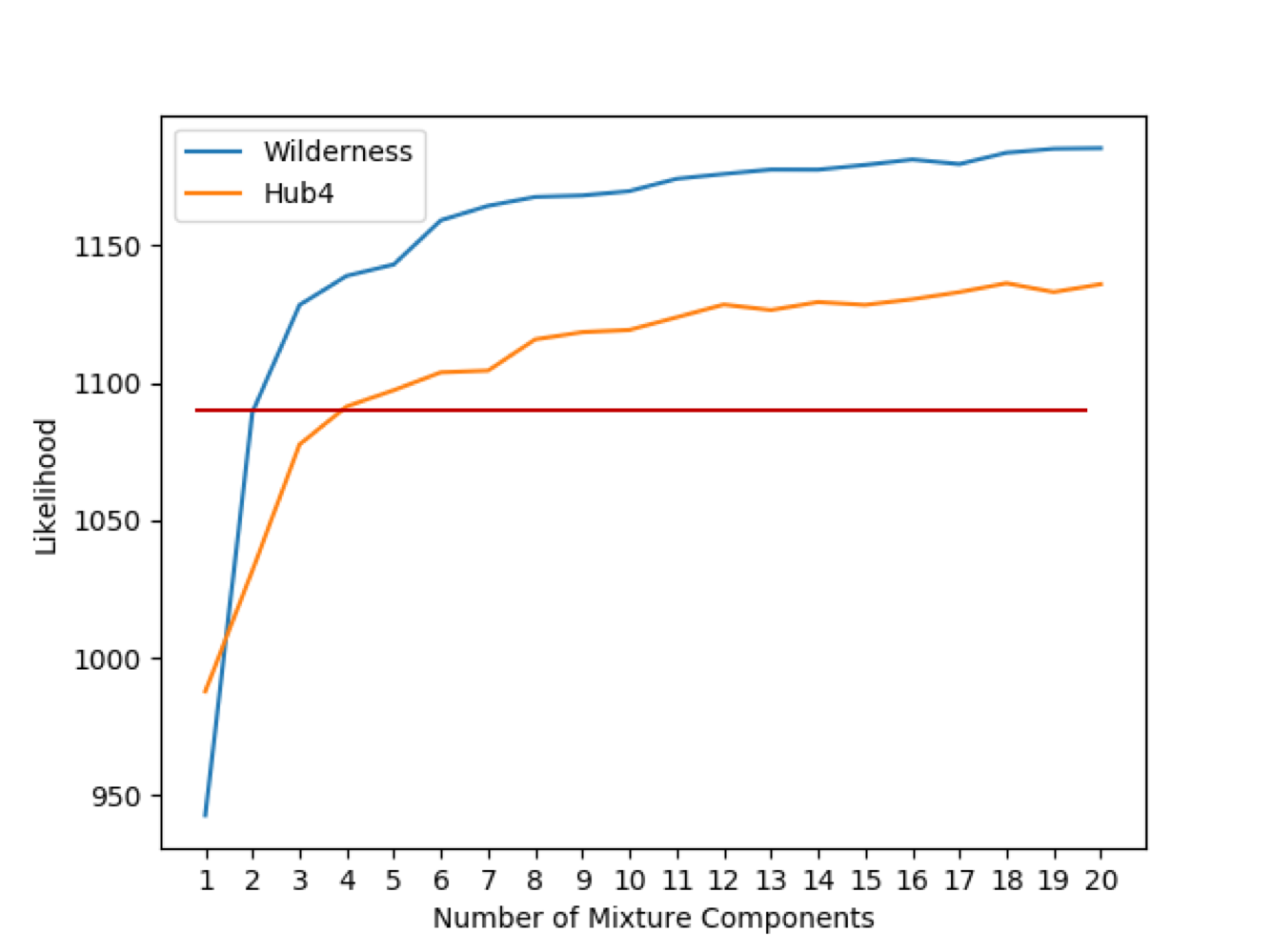}
    \caption{Gaussian Mixture Fit for Wilderness and Hub4}
    \label{fig:GMM}
\end{figure}

Figure \ref{fig:GMM} shows the likelihood of fitting Gaussian Mixture Models as a function of the number of cluster centers. As discussed earlier, speech modes are dominant in the target data so fitting $n$ clusters in the curve can be thought of as having $n-1$ nodes/clusters in the VAE for the speech and $1$ cluster as the residual non-speech data. The multinode VAE model for the wilderness data obtained good results with just $1$ VAE node or $2$ clusters as can be confirmed from the graph where likelihood values are high for just $2$ clusters. On the other hand, multinode VAE model for Hub4 gave good results with $3$ nodes or $4$ clusters. Now, as we increase the number of nodes, the peak performance does not change much, however, we attain the same peak performance for more model states:- MSE loss vs KL loss. This will be explained using training loss curves. It would have been better to use some validation parameter but since model performance for human hearing can only be analyzed by listening to the speech, we use the training metrics. \\

\begin{figure}[h!]
    \begin{subfigure}{0.4\textwidth}
    \centering
    \includegraphics[scale=0.31]{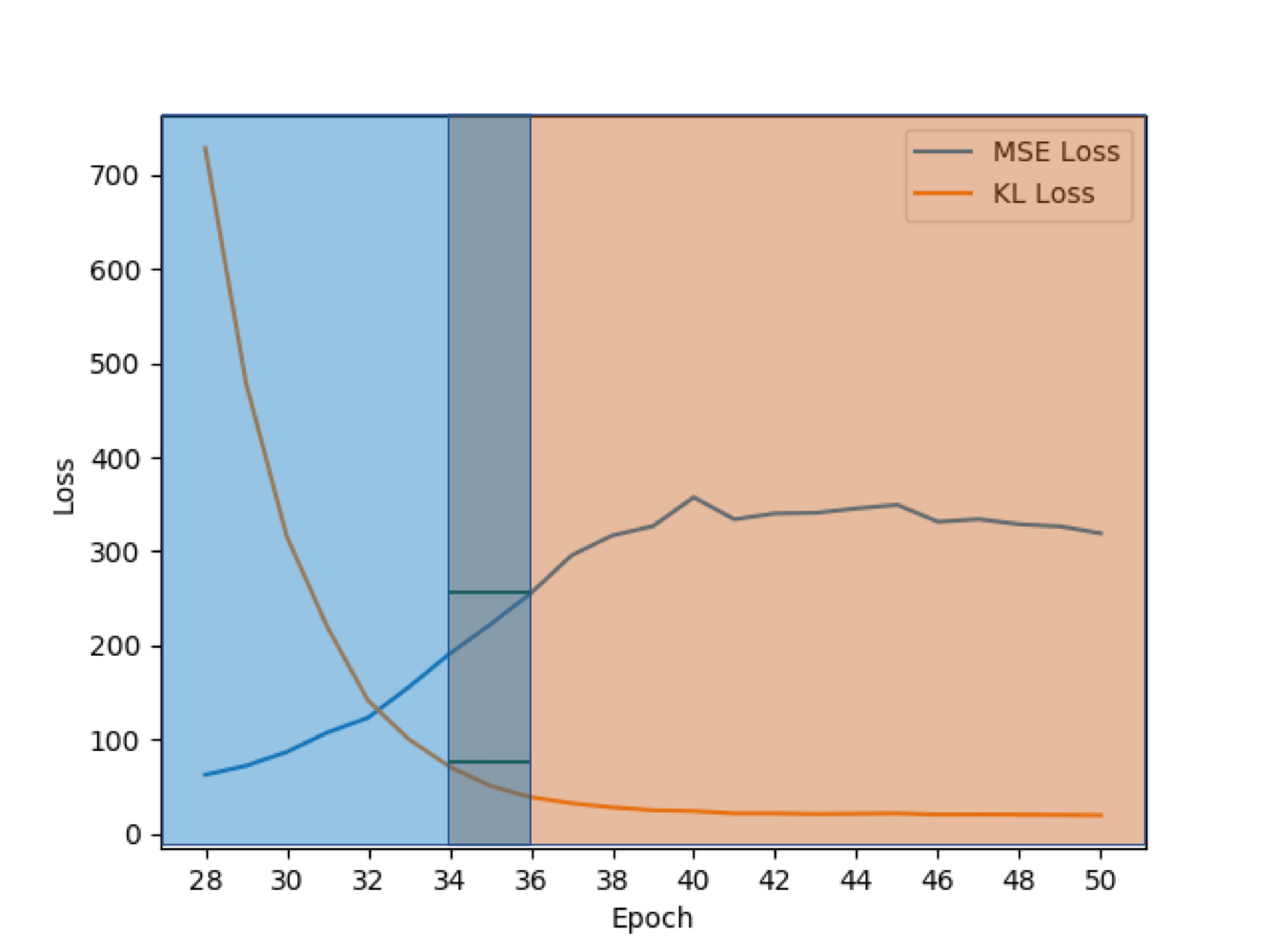}
    \caption{1-Node VAE: Wilderness}
    \end{subfigure}
    \vspace{0pt} \newline
    \begin{subfigure}{0.4\textwidth}
    \centering
    \includegraphics[scale=0.31]{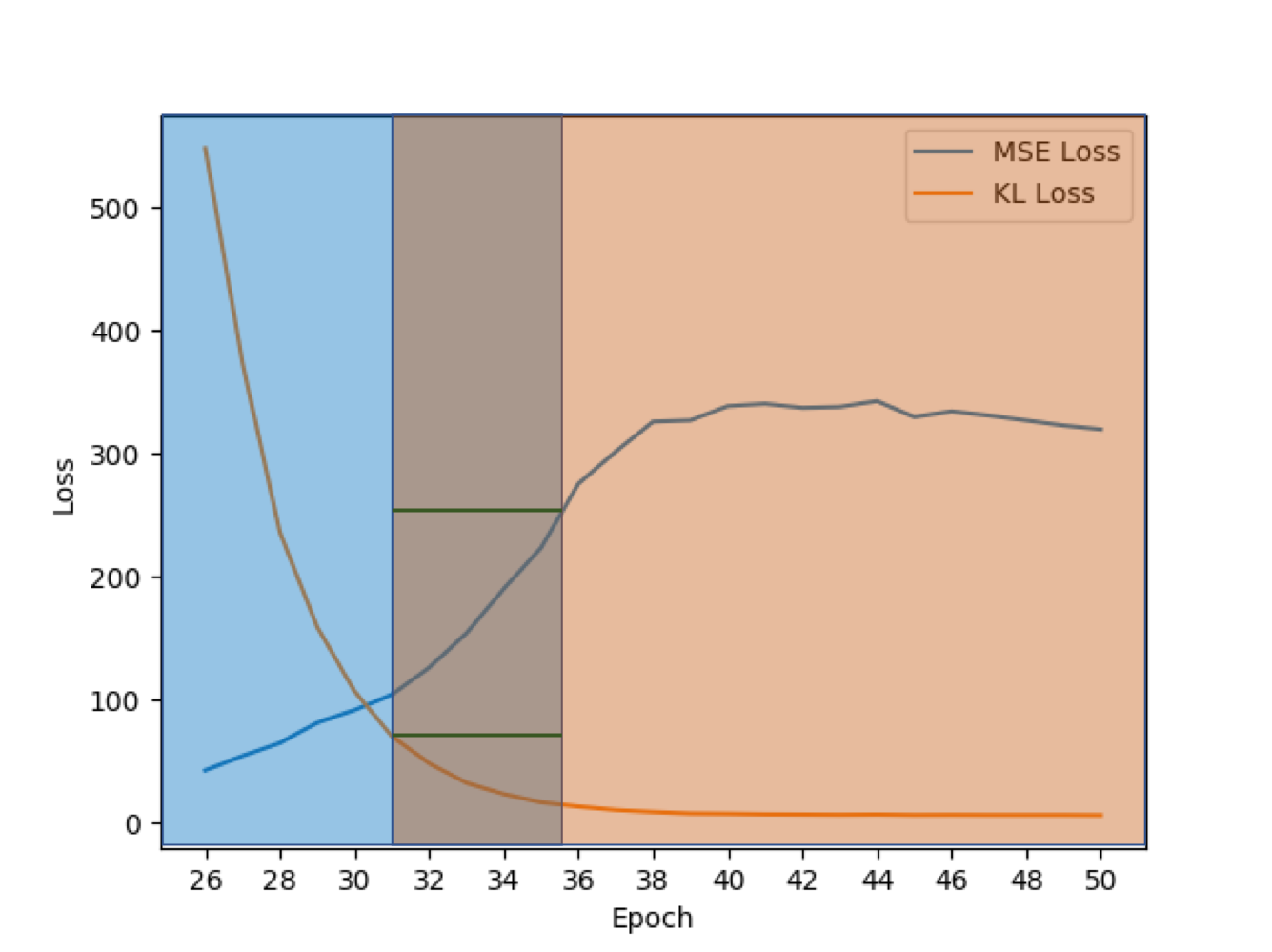}
    \caption{3-Node VAE: Wilderness}
    \end{subfigure}
    \caption{Training Loss (a) 1-Node VAE: Wilderness (b) 3-Node VAE: Wilderness. The left shaded blue region and the right shaded orange region show the required MSE loss and KL loss threshold respectively to obtain good speech. Overlap region represents the model parameters where speech separation occurs.}
    \label{fig:lossw}
\end{figure}

\begin{figure}[h!]
    \begin{subfigure}{0.4\textwidth}
    \centering
    \includegraphics[scale=0.31]{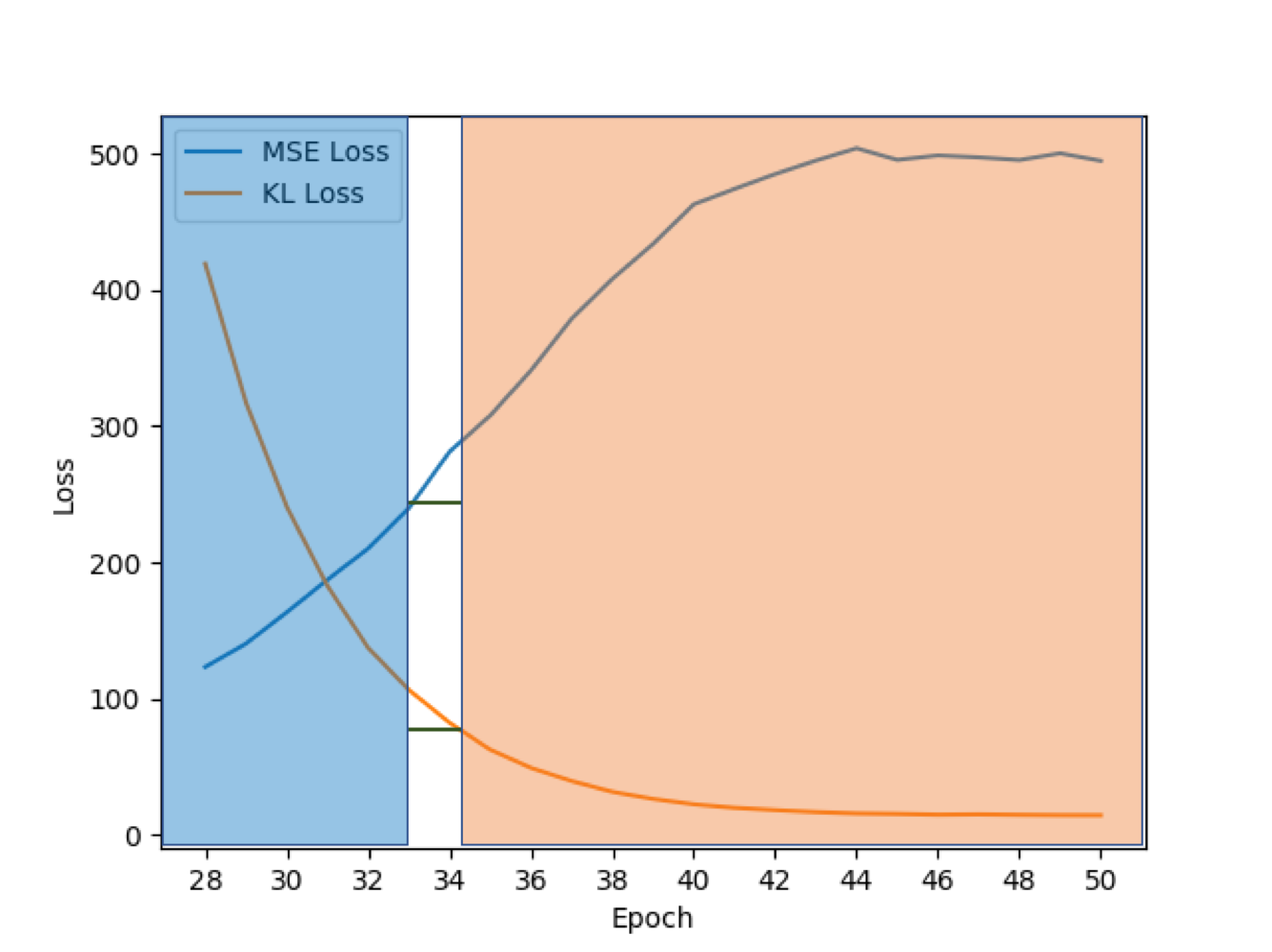}
    \caption{1-Node VAE: Hub4}
    \end{subfigure}
    \vspace{0pt} \newline
    \begin{subfigure}{0.4\textwidth}
    \centering
    \includegraphics[scale=0.31]{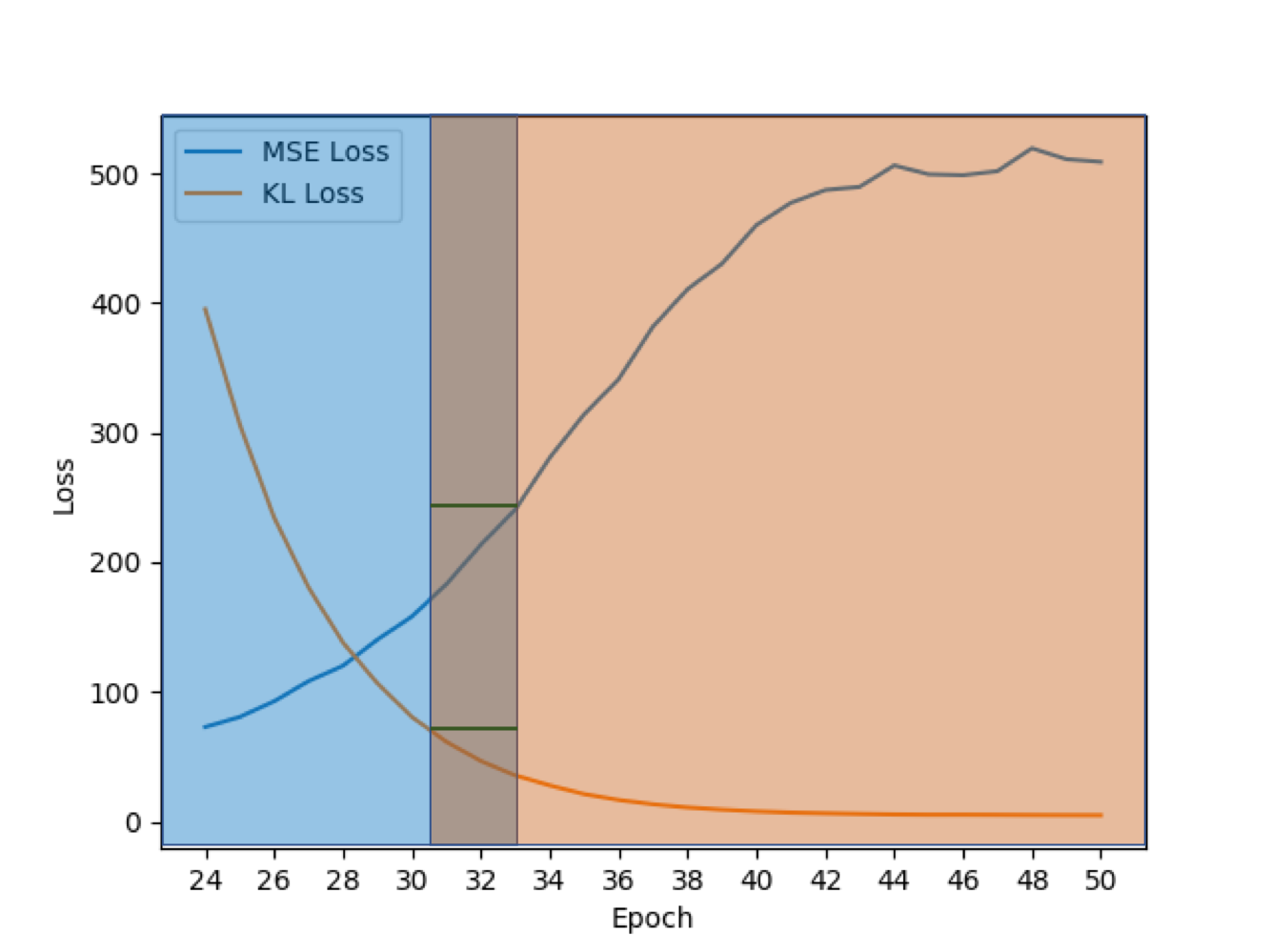}
    \caption{3-Node VAE: Hub4}
    \end{subfigure}
    \vspace{0pt} \newline
    \begin{subfigure}{0.4\textwidth}
    \centering
    \includegraphics[scale=0.31]{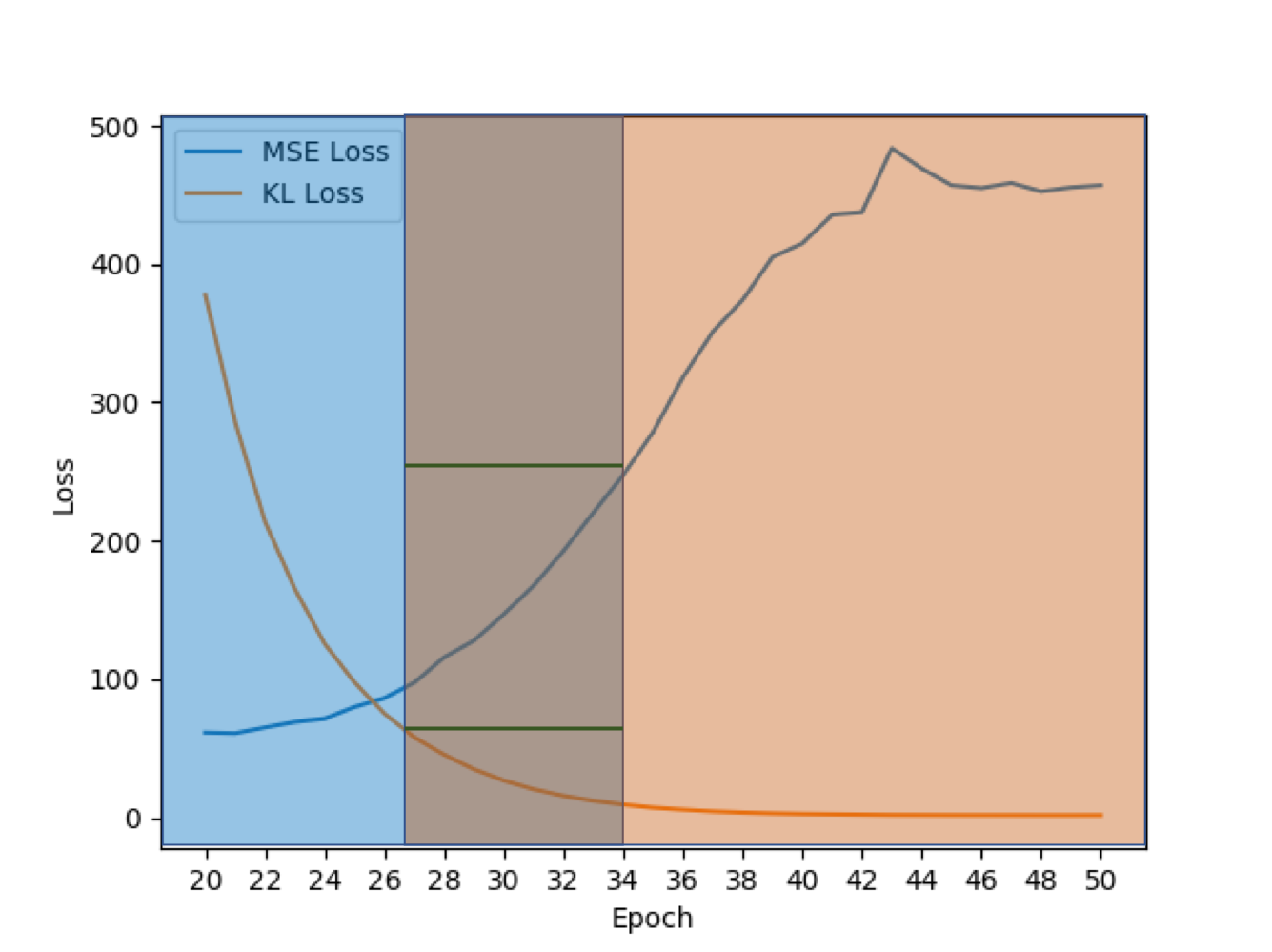}
    \caption{8-Node VAE: Hub4}
    \end{subfigure}
    \caption{Training Loss (a) 1-Node VAE: Hub4  (b) 3-Node VAE: Hub4 (c) 8-Node VAE: Hub4. The left shaded blue region and the right shaded orange region show the required MSE loss and KL loss threshold respectively to obtain good speech. Overlap region represents the model parameters where speech separation occurs.}
    \label{fig:lossh}
\end{figure}

The Figure \ref{fig:lossw} and \ref{fig:lossh} give an idea of speech separation capacity of the model as we increase the number of latent variables. During training of the multinode VAE model, we anneal the KL divergence loss for latent space exponentially. We do the annealing for latent variables simultaneously. Initially, KL divergence loss is assigned a very small weight and then increased exponentially. So, it first increases (not shown in the plot as it's out of the range of the plot) and then decreases eventually while the MSE reconstruction loss first decreases and then increases slightly. During this process there is a small window where both the losses are low enough and we are able to extract out speech from the audio. This window is determined by the threshold values for both losses. If both loss values are below their respective thresholds, we observe speech at the output.\\

The blue and orange shaded regions in figures \ref{fig:lossw} and \ref{fig:lossh} depict the loss values below the threshold for MSE loss and KL divergence loss respectively. Therefore, their intersection as indicated by the overlap region represents the model parameters that result in speech and music separation. For visualization, the MSE loss in the figure is averaged over all the samples as well as in the time dimension of the audio while the KL divergence loss is averaged over all the latent variables as well as across all samples. Using these loss definitions results in a threshold value of 250 for MSE loss and a threshold value of 60 for KL divergence loss for both datasets. These are just soft experimental values and may change for other datasets as well as a different loss definition. The key idea is that there exists a window where speech separation occurs. \\

As shown in the figure, for Wilderness data we obtain this window with just one node in the latent space. As we increase the number of nodes in the latent space, we don't see any significant improvement in the quality of the output speech, however, we do obtain a wider window where this separation occurs. As for the Hub4 dataset, we don't observe any such window with one latent node, however, we do obtain a separation window with three latent nodes and an even wider window with eight latent nodes. Observe that, these results align with the results derived using input data distribution and GMM fitting analysis. Therefore, to be totally certain about the existence of a separation window, we can always add a few more latent variables than what we obtain from our analysis of input distribution. \\

Let's explore the nature of the output on either side of the separation window. On the blue/left side of the window MSE loss is very low while the KL divergence loss is high, this results in the output that is close to the original input that consists of both speech and music. On the orange/right side of the window, MSE loss is high while the KL divergence loss is low. This causes the network output to be a really noisy version of the speech component of the audio.  \\

We also observed that, as the intensity/loudness of music in the background increases in an audio or for a part of the audio, the speech separation performance for that part of the audio begins to deteriorate slightly. For example, in case of advertisement segments between news broadcasts where music tends to dominate the segment. In such cases, we can still hear some traces of music in the background when we use the same VAE model as we do for the rest of the data. However, this is not a concern as the downstream applications we target don't generally depend on data with such high intensity music. \\

As mentioned in \cite{JMLR:v19:17-704}, a traditional autoencoder is not able to perform the outlier removal. We verified this fact experimentally. We removed any constraint on the latent space and trained the model to minimize the reconstruction loss. We observed that the model was able to reconstruct the audio completely including both speech and music. Therefore, a traditional autoencoder is not able to smooth out the energy contour and fails to remove any outliers. We also tried to experiment with this model on songs and movie clips. As the background music in songs and movies very dense and varies significantly, we observed that our model wasn't able to separate out speech completely. The output speech contained some music in the background and the quality of speech itself was compromised.  


\section{Results}
The separated speech samples for both Wilderness and Hub4 datasets can be found at \footnote[1]{https://github.com/nishantgurunath/source\_separation/tree/master/samples}
under folders "Wilderness" and "Hub4" respectively. We present Wilderness samples for two languages - Dhopadhola and Marathi both having different but somewhat uniform music in the background. These samples can be found inside their respective folders within the wilderness folder. We also present samples for when we removed the KL-divergence term from the loss function and trained an autoencoder instead. It can be observed from the samples that the model reconstructs the whole audio without outlier/music removal. Hub4 data samples have a lot more variation in both speech and music. All the samples come from news broadcast in English. We also ran experiments on mixed signals where a background music was added to a clean sample from the wilderness data. We performed this experiment with drums, flute, guitar and piano music in the background. 
The mixed audio and the separated speech samples can be found at \textsuperscript{\rm 1}.\\ 

One other way to asses the proficiency of the proposed method is to evaluate its performance on downstream tasks, for instance, Text-2-Speech Synthesis (TTS). We performed Text-2-Speech synthesis on original and cleaned version of Marathi language from the wilderness dataset. 
The TTS samples can be found at \textsuperscript{\rm 1} as well.
We observe that Text-2-Speech synthesis performance improves significantly after music is removed from the background. In TTS samples generated from the original (noisy) version of the audio, one can clearly see the distortion in the speech and presence of music in the background. Whereas, TTS samples that were generated from cleaned samples had a very clear speech quality with no music in the background. 


\section{Conclusion}
We show that Multinode VAE model helps to remove the background noise/music in the 'found data' irrespective of the language of the speech. Extensive studies on different type of speech and music data verify the effectiveness and robustness of the proposed approach. Performance of this model on Text-2-Speech synthesis applications shows the potential of such an approach that can be further extended to other speech based machine learning models such as Automatic Speech Recognition (ASR). Such an efficient source separation technique can help overcome a major cause for under utilization of 'found data'. This could mean that acoustic based machine learning models can be drastically improved by leveraging the data found on the internet. Since this approach works in a unsupervised fashion, it eliminates the need to obtain labeled data which has been major hindrance to effective utilization of 'found data'. Since 'found data' is abundant, this could also possibly further accelerate the research in this area.

\section{ Acknowledgments}
The authors appreciate the effort of the people involved in the compilation of the datasets.

\bibliographystyle{aaai}
\bibliography{bibfile}
\end{document}